# A Random Walk Model for Item Recommendation in Folksonomies

**Zhu Zhang**[1], **Daniel Zeng**[1,2], **Ahmed Abbasi**[3], **Jing Peng**[4]

[1]Institute of Automation, Chinese Academy of Sciences

[2]Department of Management Information Systems, The University of Arizona

[3]McIntire School of Commerce, University of Virginia

[4]Department of OPIM, The Wharton School, University of Pennsylvania

zhu.zhang@ia.ac.cn, dajun.zeng@ ia.ac.cn, abbasi@comm.virginia.edu, jingpeng@wharton.upenn.edu

**Abstract** *social tagging, as a novel approach to information organization and discovery, has been widely adopted in many Web2.0 applications. The tags provide a new type of information that can be exploited by recommender systems. Nevertheless, the sparsity of ternary <user, tag, item> interaction data limits the performance of tag-based collaborative filtering. This paper proposes a random-walk-based algorithm to deal with the sparsity problem in social tagging data, which captures the potential transitive associations between users and items through their interaction with tags. In particular, two smoothing strategies are presented from both the user-centric and item-centric perspectives. Experiments on real-world data sets empirically demonstrate the efficacy of the proposed algorithm.*

## 1. Introduction

In recent years, social tagging has become increasingly popular in many Web2.0 applications, including social bookmarking sites (e.g., del.icio.us, CiteULike), music sharing sites (e.g., last.fm), E-commerce sites (e.g., Amazon), etc. Social tagging systems allow users to annotate resources (items) with descriptive words of their own choice, providing a novel way for users to organize Web resources of interest and discover new resources and friends. The semantic embedded in tags represents the preference of users and the topical concern of items. These tags provide an interesting opportunity to research in recommender systems they constitute an additional source of information about the interaction between users and items in transactional database [1].

However, an important problem that remains to be studied in tag-based recommendation is the data sparsity, which has been reported to limit the performance of recommender systems [2]. In traditional transactional datasets, data sparsity appears in the form of user-item interactions. Whereas, in social tagging systems, the sparsity is shown in the form of three dimensional user-item-tag interaction. Most of the previous work pertaining to tag based recommendation focused on how to incorporating tagging information when analyzing the user-item interactions [1, 3, 4]. Nevertheless, only few studies explored the sparsity problem of item recommendation in social tagging systems [3]. In this paper, we present a novel random-walk-based recommendation algorithm which utilizes a probability-based method to compute the inter-user and inter-item similarities using item-tag, user-tag, and user-item co-occurrence information.

Experimental results on three real-world datasets show the proposed approach can efficiently alleviate the sparsity problem and improve the quality of tag-based recommendation, as compared to several benchmark methods.

## 2. Related Work

Prior work related to tag based recommendation mostly focus on how to best incorporate social tags into recommendation algorithms. For instance, Tso-Sutter et al. [1] extended user and item profiles to include user and item tags; the resulting user-based and item-based methods were combined into a fusion method. Wetzker et al. [4] extended the probabilistic Latent Semantic Analysis (PLSA) approach and presented a recommendation model which estimated from item-user and item-tag observations in parallel. Peng et al. [5] presented a joint item-tag recommendation framework, which explicitly pointed out the topical interests of users in the recommended items and made full use of all available interactions among users, items, and tags. However, very little research in the literature has studied the sparsity problem in the context of social tagging systems. As a result, this paper aims at dealing with sparse problem of social tagging data.

## 3. Random-Walk-based Item Recommendation

### 3.1 Notations

Let $U = \{u_1, u_2, u_3, \cdots, u_m\}$ be the set of users, $I = \{i_1, i_2, i_3, \cdots, i_n\}$ be the set of items, and $T = \{t_1, t_2, t_3, \cdots, t\}$ be the set of tags. We use UT, UI, and IT to represent user-item, user-tag, and item-tag co-occurrence matrices, respectively. If user $i$ saves item $j$, the element $UI_{ij}$ in the matrix UI is one, and otherwise zero. An entry $UT_{ik}$ in matrix UT represents the frequency of tag $k$ used by user $i$, and an entry $IT_{jk}$ in matrix IT represents the frequency of tag $k$ labeled to item $j$. $M_{norm}$ is a matrix generated by normalizing each row of M to unit sum. $UI_{i\cdot}$ indicates the $i$th row vector of UI, and $UI_{\cdot j}$ indicates the $j$th column vector of UI.

### 3.2 Overview of random walk based method

A random walk over a graph is a stochastic process in which the initial state is known and the next state is governed by a transition probability matrix that dictates the likelihood of jumping from node $i$ to node $j$ in the graph [6, 7]. The random walk method has been used in prior studies to capture the implicit transitive associations between (item) graph nodes, thereby facilitating the alleviation of sparsity problem and the accurate recommendation of top-ranked items [2, 6, 7]. We propose a novel random walk based recommendation algorithm specifically designed to harness the unique characteristics of social tagging systems for enhanced item recommendation. The novel contributions of our proposed method (over previous work) include the following. In our approach, we incorporate item-tag and user-item interaction information in the item graph. Unlike prior work [2, 6, 7], we use the enriched item graph in conjunction with a user graph comprised of user nodes and edges that signify the similarity between users. We also present a novel probabilistic method for incorporating tag information into the computation of the item and user similarity matrices. Collectively, these facets of the proposed method are intended to facilitate enhanced item recommendation, even when encountering sparse social tagging data. As later demonstrated in the evaluation section, each of these aspects contributes to its enhanced performance over existing methods. Details of our method are presented in the remainder of the section.

### 3.3 Computation of inter-item and inter-user similarities

A critical step to build a random walk model is to compute similarities. In this paper, we use item and user similarity matrices as transition probability matrices, because we argue that the more is item *i* similar with item *j*, the higher is the probability that a random walker jumps from item *i* to item *j*. During the computation of item similarity using UI, Deshpande et al. [8] empirically shown it would improve the recommendation performance to assign more weight to the purchase decision of customers that bought few items by normalizing UI. Inspired by this work, we propose a novel probability based similarity method for deriving the item similarity method, which has a good interpretation based on transition probability. The specific is shown below.

$$S_{item} = \alpha \cdot IT_{norm} \cdot IT_{norm}^T + (1 - \alpha) \cdot UI_{norm}^T \cdot UI_{norm} \quad (1)$$

In the above equation, we incorporate IT and UI into the calculation of item similarity matrix, because IT contains the semantic content information of items and UI contains the user-item interaction information, the combination can make the representation of item similarity more precise. The rationales behind both methods are identical, so we just take $IT_{norm} \cdot IT_{norm}^T$ as an example. We can get the probability from one item to all the tags from $IT_{norm}$ and the probability from one tag to all the items from $IT_{norm}^T$. Then, the similarity between item *i* and item *j* can be computed as the dot-product of the *i*th row vector of $IT_{norm}$ and the *j*th column vector of $IT_{norm}^T$, which represents the probability that item *i* jumps to item *j* through all of the tags. Likewise, we can compute the user similarity matrix as follows:

$$S_{user} = \beta \cdot UT_{norm} \cdot UT_{norm}^T + (1 - \beta) \cdot UI_{norm} \cdot UI_{norm}^T \quad (2)$$

### 3.4 Random walk based recommendation algorithm

We firstly use item similarity matrix as the transition probability matrix, and each user takes a random walk on the item graph. For each user, this results in an item vector that signifies that user's predicted probabilities for the items. The basic idea is that each user will choose new items similar to the ones they have chosen in the past. Similarly, we use the user similarity matrix as transition probability matrix, and each item takes a random walk on the user graph. For each item, this results in a user vector that can predict the probability of the different users' choices for that item. Here, the underlying idea is that users will choose items that were selected by similar users. According to the above principle, our recommendation algorithm based on random walk is as follows:

$$UI^{item}(0) = UI^{user}(0) = UI_{norm} \quad (3)$$

$$UI_{i\cdot}^{item}(t+1) = \eta \cdot UI_{i\cdot}^{item}(t) \cdot S_{item} + (1-\eta) \cdot UI_{i\cdot} \quad (4)$$

$$UI_{\cdot j}^{user}(t+1) = \lambda \cdot S_{user} \cdot UI_{\cdot j}^{user}(t) + (1-\lambda) \cdot UI_{\cdot j} \quad (5)$$

$$UI_{final} = \mu \cdot UI_{final}^{item} + (1-\mu) \cdot UI_{final}^{user} \quad (6)$$

where $UI^{item}$ and $UI^{user}$ are the item-centric and user-centric prediction scores, respectively. The parameters $\eta$ and $\lambda$ are the damping factors. Larger values for these parameters increase the importance of the hidden connections captured by the multi-step random walks. However, the iterations to convergence increase as the two damping factors increase. Every row vector of UI normalized in the Eq. (4) and (5) is the preference vector of the corresponding user. When $UI^{item}(t)$ and $UI^{user}(t)$ converge to the acceptable optimal performance, we can get the final user-item predicted score matrix $UI_{final}$ by fusing them. In particular, as the number of iterations goes to infinity, Eq. (4) can be reduced to

$$UI^{item} = (1-\eta)UI \sum_{k=0}^{\infty}(\eta S^{item})^k = (1-\eta)UI(1-\eta S^{item})^{-1} \quad (7)$$

$$UI^{user} = (1-\lambda)\sum_{k=0}^{\infty}(\lambda S^{user})^k UI = (1-\lambda)(1-\lambda S^{user})^{-1}UI \quad (8)$$

The algorithm converges to the optimal performance in roughly 5 iterations in our experiments. From Eq. (7) and (8), we can see our approach is the fusion of item based and user based CF in nature. The reason for this method to address the sparsity problem is that capturing the hidden transitive associations among the user-tag-item tripartite graph by multi-step random walks makes the item and user similarity matrices denser and more accurate compared with traditional item-based and user-based CF.

## 4. Empirical study

We evaluated the proposed algorithm using three tagging data sets. During data preprocessing, we iteratively removed users that had saved less than 10 items and items that had been saved by less than 10 users (8 for the Bibsonomy data set), until the number of unqualified items were less than 20 for each data set. Table 1 shows the key statistics of the cleaned data sets. For each user, we randomly selected 20% of the items for training and withheld the rest for prediction. Of these user test items, we recommended the top 5 items to each user.

**Table 1.** Data set description

| Data set | CiteULike | Bibsonomy | Delicious |
|---|---|---|---|
| Number of users: m | 338 | 125 | 177 |
| Number of items: n | 392 | 388 | 210 |
| Number of selected/total tags: l | 24/2822 | 84/2311 | 68/2251 |
| Number of total transactions: p | 6031 | 4883 | 4093 |
| Data density: p/(mn) (%) | 4.55 | 9.04 | 11.01 |
| Avg. number of items per user | 17.84 | 35.06 | 23.12 |
| Avg. number of users per item | 15.39 | 11.30 | 19.49 |

The evaluation metrics used were the commonly used ones in the literature [5], including precision, recall, F-measure, and rankscore for ranked list prediction. We compared the proposed approach with four other approaches. Two of them were the classical user-based and item-based methods, and the other two were tag-based recommendation methods: the fusion [1] and PLSA [4] methods. pRW-IT only used the IT matrix, and pRW-UT only used the UT matrix. pRW-UI only utilized the UI matrix. In contrast, pRW used both the IT and UI matrices. The reported results were the averages over ten different runs.

Table 2. Top 5 recommendation on CiteULike, Bibsonomy and Delicious(in order)

| Alg.\metrics | Precision (%) | | | Recall (%) | | | F-measure (%) | | | Rankscore (%) | | |
|---|---|---|---|---|---|---|---|---|---|---|---|---|
| Random | 4.05 | 6.82 | 9.27 | 1.36 | 1.19 | 2.52 | 2.04 | 2.02 | 3.96 | 4.03 | 6.81 | 9.15 |
| User based | 10.99 | 13.90 | 25.48 | 4.63 | 2.84 | 7.12 | 6.51 | 4.72 | 11.13 | 11.39 | 14.20 | 26.04 |
| Item based | 7.67 | 10.70 | 18.38 | 3.43 | 2.34 | 4.87 | 4.47 | 7.69 | 7.63 | 7.80 | 10.87 | 18.40 |
| Fusion | 12.67 | 18.88 | 29.63 | 5.28 | 4.15 | 8.64 | 7.45 | 6.81 | 13.38 | 12.80 | 18.96 | 30.31 |
| PLSA | 12.72 | 15.49 | 29.23 | 5.23 | 3.07 | 8.17 | 7.41 | 5.12 | 12.76 | 13.02 | 15.71 | 29.91 |
| pRW-IT | 14.49 | 18.86 | **31.75** | 5.93 | 3.93 | **9.17** | 8.41 | 6.50 | **14.23** | 14.78 | 19.36 | **32.67** |
| pRW-UT | 10.68 | 16.66 | 27.04 | 4.24 | 3.25 | 8.03 | 6.07 | 5.44 | 12.38 | 11.01 | 17.00 | 27.82 |
| pRW-UI | 10.70 | 16.51 | 28.49 | 4.56 | 3.42 | 8.15 | 6.39 | 5.67 | 12.67 | 11.10 | 16.89 | 29.28 |
| pRW | **15.18** | **19.79** | **31.75** | **6.32** | **4.26** | **9.17** | **8.91** | **7.01** | **14.23** | **15.55** | **20.36** | **32.67** |

As shown in Table 2, pRW-IT outperformed pRW-UI, demonstrating that the tagging information was more effective than the transactional information in the computation of item similarity. However, the results of pRW-UT and pRW-UI are similar, implying that the tagging information doesn't outperform the transaction information in the computation of user similarity. The use of random walks in conjunction with the application of the probabilistic similarity based mechanism on the IT, UT and UI matrices enable pRW to outperform all comparison methods on all of evaluation conditions depicted in Table 2. According to t-test, pRW is significantly better than the second best algorithm Fusion, with $p<0.002$ on all evaluation metrics for both CiteULike and Delicious datasets.

In order to evaluate the performance of our approach under sparse data, we conducted several experiments on the CiteULike data set at different density levels. From Table 3, we can see that pRW outperformed the other algorithms under sparse data (the results of other metrics is similar and are omitted due to the space limit). According to t-test, pRW is significantly better than the second best algorithm PLSA, with $p<0.01$ on all evaluation metrics when training set percentage is 10% and 20%. The reason for this was that the multi-step random walks were able to capture the hidden transitive relation among users, tags, and items that were ignored in the other methods.

Table 3. Precisions of top 5 recommendations on the CiteULike

| Algorithm | Training set percentage | | |
|---|---|---|---|
| | **5%** | **10%** | **20%** |
| User based | 4.86 | 7.41 | 10.99 |
| Item based | 4.92 | 6.91 | 7.67 |
| Fusion | 7.08 | 9.54 | 12.67 |
| PLSA | 7.23 | 10.01 | 12.72 |
| **pRW** | **7.57** | **10.81** | **15.18** |

## 5. Conclusion and discussion

We proposed a novel random-walk-based recommendation algorithm that can effectively capture the transitive associations between users and items to deal with the sparsity problem in tag-based recommendation. The experiments empirically demonstrate the superiority of the proposed algorithm over existing tag-based recommendation methods. For future work, we plan to consider the polysemy and synonyms of tags, e.g., clustering tags based may further improve the performance of our method. In addition, we also plan to incorporate social network information into our recommendation framework in the future.

**Acknowledgement.** This research is partially funded through CAS Grant #2F07C01, NNSFC Grants #71025001, #91024030, #90924302, #70890084, and#60875049.